\documentstyle[epsfig]{elsart}

\begin{document}

\vspace*{2cm}
\begin{frontmatter}
 
\title{Simulation studies of the HADES first level trigger.\\
PART I: Performance in heavy-ion induced reactions}

\author{R.~Schicker\thanksref{corr}}
\author{and H.~Tsertos}
\address{University of Cyprus, Nicosia, Cyprus}

\begin{abstract}

The first level trigger of the HADES spectrometer is studied for the heavy-ion 
collision systems Au+Au and Ne+Ne. The trigger efficiency for central events 
is given in dependence of the imposed charged particle multiplicity condition. 
The timing properties of the trigger signal are described. The losses due to 
deadtime are specified. Finally, the first level trigger rate is reported.

\end{abstract}

\thanks[corr]{Corresponding author, e-mail "schicker@alpha2.ns.ucy.ac.cy" \\
Dept. Nat. Science, Univ. Cyprus, PO 537, 1678 Nicosia, Cyprus}

\end{frontmatter}

\vspace{-13.cm}
\hspace{10.cm}{\bf UCY--PHY--96/14}
\vspace{12.0cm}

\newpage

\section{Introduction} 
\label{sec:intro}

The dilepton spectrometer HADES is currently being built at the heavy-ion 
synchrotron SIS at GSI Darmstadt\cite{hades1}. HADES will measure dielectron 
pairs emitted in relativistic heavy-ion collisions in the beam energy range of 
1-2 AGeV. Additionally, a secondary pion beam facility of momenta between 0.5 
GeV/c and 2.5 GeV/c will allow the measurements of dilepton observables in pion 
induced reactions\cite{pion}. Measurements of in-medium dielectron decays of 
the vector mesons $\rho,\omega$ and $\phi$ allow to reconstruct in-medium 
masses of these vector mesons. Hence, experiments with HADES will test a 
series of conjectures about in-medium behavior of vector mesons in 
hot and dense matter produced in relativistic heavy-ion collisions\cite{Brown}.

The HADES spectrometer consists of a superconducting toroidal magnet with six 
sectors of 60$^{0}$ azimuthal angle each. The acceptance in polar angle extends 
from 18$^{0}$ to 85$^{0}$ in the forward hemisphere. Multiwire Drift Chambers 
(MDCs) placed forward of and behind the magnetic field are used for trajectory 
reconstruction of the individual particles. Between the target and magnetic 
field region, a Ring Imaging Cherenkov detector (RICH) identifies electron 
trajectories with a threshold $\gamma_{thr} \sim$ 20. In the region behind the 
magnetic field, a second, independent electron identification is achieved by a 
Multiplicity/Electron Trigger Array (META). The META system consists of 
time of flight (TOF) paddles and shower detectors\cite{hades1}.

HADES is designed to operate at heavy-ion beam intensities of 10$^{8}$ particles
per second. A 1\% interaction target results in 10$^{6}$ minimum bias events per 
second. The multi-level trigger system of HADES has to reduce this primary rate 
to a few hundred events per second which will be written onto tape. 
The first level trigger is designed to select central collisions by recognizing 
events with a large charged particle multiplicity. The expected rate of the 
first level trigger is 10$^{5}$ events per second. The second level trigger 
searches for lepton candidates by matching the identified RICH rings with the 
measured META showers. This matching 
procedure is based on the angular correlations of physical tracks in the RICH 
and META detectors. The third level trigger combines the output of the matching 
unit with the tracking information from the MDCs. This trigger stage rejects 
mainly combinations of low energy electrons producing a RICH ring with hadrons 
misidentified as leptons in the META system.

The purpose of this paper is to present simulation studies of the HADES first 
level trigger in heavy-ion induced reactions. Simulation results are shown for 
the systems Au+Au at 1 AGeV and Ne+Ne at 2 AGeV, in order to illustrate the 
trigger performance in a heavy and a light collision system, respectively.

This paper is organized as follows: Section \ref{sec:ftrig} gives a summary 
on the HADES first level trigger requirements. In Section \ref{sec:pdat}, the 
simulation of the first level trigger data is described. Section 
\ref{sec:adat} introduces the analysis of the simulation data. In Section 
\ref{sec:fauau}, the performance characteristics of the first level trigger 
in the heavy system Au+Au is presented. Section \ref{sec:fnene} gives details 
of the first level trigger performance in the light system Ne+Ne.

\section{First level trigger requirements}
\label{sec:ftrig}

The HADES first level trigger will tag central heavy-ion reactions 
by recognizing events with a large charged particle multiplicity. This 
multiplicity can be derived from the highly segmented TOF array. A lower limit 
on the number of TOF paddles carrying a signal will select central events. 
An optimal first level trigger accepts all central events while suppressing 
non-central events as much as possible.

Comprehensive dilepton spectroscopy of heavy-ion reactions, however, 
necessitates additionally the measurement of the dielectron signal in 
non-central events. With the same first level trigger approach as described 
below, data of non-central events can be taken by redefining the multiplicity
conditions and by appropriately downscaling the resulting trigger rate. Thus, 
non-central events can either be downscaled and be registered simultaneously
with central events or can be recorded in dedicated data taking periods.

The trigger signal derived from the multiplicity conditions of the TOF paddles 
is used as gate for the ADCs of the RICH detector and as common STOP for the 
TDCs. Hence the delay of this signal with respect to the time of reaction as 
well as the time jitter are of particular interest. The time jitter of the first
level trigger signal arises from different sources. First, trajectory length 
variations over the polar angular range of the spectrometer induce particle 
time of flight variations. Second, velocity variations of the particles 
defining the trigger transition add to the time of flight variations. Third, the
different signal propagations in the TOF paddles depending on location of the 
hit point add varying delays to the TOF signals.

Each signal of the first level trigger has an associated deadtime due to 
frontend readout of detectors. This deadtime is about 10 $\mu$sec. 
If the multiplicity condition is set low, then the trigger rate will be 
increased with a corresponding increase in probability that central events 
fall into the deadtime window. If, on the other hand, the multiplicity 
condition is set high, then central events start to get rejected due 
to insufficient multiplicity. For a given deadtime, there exists 
therefore a multiplicity condition which optimizes the number of central 
events which are passed onto the next trigger stage.

Due to the statistical occurrence of heavy-ion reactions, there is a finite
probability that two or more events occur very close in time.
Hence, TOF paddles can carry simultaneously signals of different
events. The combined signals of the different events may satisfy the 
trigger condition whereas none of the individual events would be able 
to do so. This overlap probability depends strongly on
the reaction rate and, thus, on beam intensity. For all of the results
shown, a beam rate of 10$^{8}$ per second and a minimum bias event
rate of 10$^{6}$ per second is assumed.

The performance of the first level trigger in heavy-ion collisions depends weakly 
on the duration of the TOF paddle signals\cite{UCY1,UCY2}. If the TOF signal 
length is short, the signals from the fastest particles will have disappeared,
while the signals from the slower particles of the event have not started yet. 
The trigger system will therefore see an apparently reduced event multiplicity. 
TOF signals which are long, on the other hand, result in an increased number 
of triggers from events overlapping in time. In this report, all the results 
shown have been derived with a TOF signal duration of 15 nsec. 

\section{First level trigger data simulation}
\label{sec:pdat}

For studying the behavior of the first level trigger, the full HADES geometry
was implemented into the GEANT package\cite{Heike}. A realistic field map of 
the toroidal magnetic field is used for tracking of the charged particles.

The collisions of the two systems studied Au+Au and Ne+Ne are modeled
by a transport equation of the Boltzmann-Uehling-Uhlenbeck (BUU) type.
The dynamical evolution of the collisions is determined by calculating
the phase space evolution for nucleons, Delta and N$^{*}$ resonances.
With this code, good agreement is found between data and model predictions
for nucleon, pion, kaon and dilepton distributions in heavy-ion collisions 
in the energy range 1-2 AGeV\cite{Wolf}. 
Since the charged particle multiplicity in central collisions is mainly
due to protons and pions, only these two particle species are tracked
for the first level trigger simulations.

For the simulations of the first level trigger in the systems Au+Au and Ne+Ne, 
BUU events of different impact parameters are used. In these systems, the upper 
limit of the impact parameter range is defined by the geometrical cross section.
In each of the two systems studied, six to eight discrete equidistant impact 
parameters represent the full range from zero to maximum value. Each tracked 
particle of an event is followed through the complete HADES geometry. A 
trajectory entering a TOF paddle volume defines a TOF hit. The information of 
this hit contains the TOF paddle number and a time value. This time value
represents the sum of the particles time of flight from target to the paddle 
hit point plus the shorter of the two propagation delays of the signal
to either end of the paddle. Here, no finite resolution or signal time jitter 
is assumed. This exact time value is then converted into an integer format with 
a discretization accuracy of 10 psec and subsequently written into a CERN 
CWN-tupel. At this stage, each discrete impact parameter is simulated 
separately. Hence, separate tupel files exist for each of the discrete
impact parameters simulated.

\section{First level trigger data analysis}
\label{sec:adat}

In the analysis of the first level trigger, a stream of events is 
generated by assigning a random time interval t$_{\Delta}$ 
and a random impact parameter b to each event. 

The time interval t$_{\Delta}$ is chosen according to the probability
$p(t_{\Delta}) = \lambda \cdot \e^{-\lambda\cdot t_{\Delta}}$. This 
probability distribution represents the distribution of time intervals
between two consecutive events, with $\lambda$ being equal to the
inverse of the average time between two consecutive events. The 
absolute time of an event is determined by adding t$_{\Delta}$ to the 
absolute time of the previous event.
                               
The impact parameter b of an event is taken from a distribution which is 
zero at b\,=\,0\,fm and linearly increasing up to b\,=\,b$_{max}$. Here, 
b$_{max}$ is defined by the geometrical cross section as explained above. 
An event is subsequently read from the tupel file with the closest nearby 
simulated impact parameter. 

The analysis of the first level trigger loops over all TOF paddles in time 
steps of 1 nsec. In each loop, the sum of the TOF paddles carrying a signal 
at that particular moment is calculated. If the given trigger multiplicity 
requirement is met in a step, but not in the previous one, then a trigger
transition is induced. In the prompt trigger described below, a trigger
transition generates a trigger. In the delayed trigger described below, a 
trigger transition initiates a time window of 20 nsec during which the event 
has to satisfy an additional multiplicity requirement. If this additional 
condition is met, then a trigger is generated. Here, prompt and delayed 
trigger refer to the one and two step trigger approach as described in 
Section \ref{ssec:fmt}. Each trigger starts a deadtime window during which 
no other triggers are accepted. 

\section{First level trigger in Au+Au collisions}
\label{sec:fauau}

\subsection{Sector multiplicity condition}

In the high multiplicity heavy-ion system Au+Au, a multiplicity condition in 
each azimuthal sector will tag central events. Fig. \ref{fig:fig1} shows the 
tagging efficiency for this system as a function of the imposed sector 
multiplicity condition M$_{S}$. This condition implies that the charged 
particle multiplicity is greater or equal to M$_{S}$ in each of the six 
azimuthal sectors. Shown are the data points for events with impact parameters 
of 1,3 and 5 fm, respectively. For all the impact parameters shown in Fig. 
\ref{fig:fig1}, the efficiency exhibits a plateau of nearly 100\% at low 
sector multiplicity values but drops steeply at large multiplicity values. 
The optimal choice for the sector multiplicity condition is a value as large 
as possible but still within the plateau of the impact parameter b\,=\,1\,fm. 
Hence, the sector multiplicity condition is set to M$_{S}\geq\,8$ for the 
calculations shown below.

\subsection{Trigger timing}

Events satisfying the sector multiplicity condition M$_{S}$ generate a trigger 
transition. The delay of this signal relative to the time of reaction and the 
signal jitter are of interest. Fig. \ref{fig:fig2} shows the timing 
of the trigger transition for events with different impact parameters.
Central events are represented by the solid line. The FWHM of their time 
distribution amounts to about 2 nsec. The FWHM value for events with impact 
parameters of 3 and 5 fm are about 4 and 6 nsec, respectively. 
Semi-central events meet the trigger condition only with the help of slower 
moving particles. Hence, for increasing impact parameters, a shift of the 
centroid to larger time values as well as a broadening of the distribution 
is seen. The time zero in Fig. \ref{fig:fig2} is the time of reaction.

\subsection{Total multiplicity}

The sector multiplicity requirement defines a condition on minimum particle 
multiplicity in each of the six azimuthal sectors. A condition on minimum total 
multiplicity might further reduce non-central events while at the same time 
accepting all central events. At the moment of the trigger transition, the 
total multiplicity for central events is still building up, whereas it is 
almost exhausted for non-central events. Fig. \ref{fig:fig3} displays the 
maximum total multiplicity reached during a time window of 20 nsec following the
trigger transition. Central and non-central events which satisfy the sector 
multiplicity condition M$_{S}$ behave quite differently in Fig. \ref{fig:fig3}.
A condition on minimum total multiplicity will therefore further reduce 
non-central events.

\subsection{Timing total multiplicity}

Fig. \ref{fig:fig4} shows the time at which the maximum event multiplicity is 
reached for events with impact parameters b=1,3 and 5 fm. Here, the time zero 
is the time of the trigger transition defined by the sector multiplicity 
condition M$_{S}$. For central events, the maximum total multiplicity 
develops between 5 and 15 nsec following the trigger transition. 
Non-central events with impact parameters b\,=\,5\,fm develop their
maximum total multiplicity during a time span of about 10 nsec following
the trigger transition. Hence a time window of 15-20 nsec duration
following the trigger transition seems adequate to test for the maximum 
total event multiplicity.

\subsection{Trigger efficiency}

Each trigger transition followed by a total multiplicity larger than a required 
minimum value M$_{T}$ generates a trigger. Each trigger has an associated 
deadtime due to frontend readout of electronic channels. An event trigger 
occuring during the deadtime window of the previous trigger will not initiate 
readout and the information of the event is lost. Fig. \ref{fig:fig5} shows 
the values of Eff$_{LV1} \times$ R$_{DT}$ for central events for M$_{S} \geq 8$ 
as a function of the required total multiplicity M$_{T}$. 
Here, Eff$_{LV1}$ denotes the first level trigger efficiency, i.e.,
the efficiency for zero deadtime. R$_{DT}$ is a reduction factor which
contains the effects of the trigger deadtime. 
For the deadtime of 10 $\mu$sec, the Eff$_{LV1} \times$ R$_{DT}$ value
has a maximum at an M$_{T}$ value of about 110. At lower M$_{T}$ 
values, semi-central events do not get suppressed efficiently. Thus, 
the increased rate of semi-central events leads to a decrease of the  
Eff$_{LV1} \times$ R$_{DT}$ values of central events. For high
M$_{T}$ values, Eff$_{LV1} \times$ R$_{DT}$ decreases since central
events start to get rejected due to insufficient multiplicity.
The data for the deadtime of 6 $\mu$sec are shown for comparison.   

\subsection{Trigger rate}

Table \ref{tab:tab1} lists the trigger rates in units of 10$^{5}$ per second
for the different total multiplicity thresholds M$_{T}$ and for deadtimes of 
T$_{0}$ = 0,6 and 10 $\mu$sec, respectively. A first level trigger rate of 
1.$\times$10$^{5}$ per second is the input design value for the next trigger 
stage. Each T$_{0}$ value in Table \ref{tab:tab1} corresponds to two columns. 
The left column shows the rate of trigger transitions resulting from the sector 
multiplicity condition M$_{S} \geq 8$. The right column displays the rate 
if additionally the total multiplicity condition M$_{T}$ is required. 

Table \ref{tab:tab2} lists the partial trigger rates from events of different 
impact parameters. For the chosen multiplicity requirements, triggers from
events with impact parameters b$\geq$5\,fm are negligible and therefore not 
listed in Table \ref{tab:tab2}. Both impact parameters b\,=\,1 and 3 fm 
correspond to two columns. On the left, the partial trigger rate is shown in 
units of 10$^{5}$ per second. On the right, the partial rates are normalized 
to the total trigger rate and shown in units of percent.

\subsection{First level trigger based on M$_{T}$}
\label{ssec:fmt}

The trigger scheme studied so far divides the first level trigger into two 
consecutive steps: In a first step, the M$_{S}$ fastest particles in each 
sector define the trigger timing by establishing a trigger transition. In a 
second step, the trigger transition is asserted or rejected within a subsequent 
time window depending on whether the total multiplicity condition M$_{T}$ is 
satisfied. However, a one step trigger scheme, based exclusively on the total 
multiplicity M$_{T}$, seems also feasible since the total multiplicity condition 
M$_{T}$ is more restrictive than the sector multiplicity condition M$_{S}$.

Fig. \ref{fig:fig6} displays the trigger timing achieved by the one and two 
step trigger. The solid line shows the timing in the two step scheme with 
conditions M$_{S} \geq 8$ and M$_{T} \geq 110$ as explained above. These data 
points correspond to Fig. \ref{fig:fig2} with the additional condition of 
total multiplicity M$_{T}$. The FWHM of this distribution is about 2 nsec. The 
dashed line in Fig. \ref{fig:fig6} represents the timing in the one step 
trigger. Here, the total multiplicity condition M$_{T} \geq 110$ defines the 
trigger. As expected, the timing is delayed compared to the two step 
trigger and considerably broadened. The FWHM of this distribution is about 
6 nsec. The data shown in Fig. \ref{fig:fig6} represent the timing of all 
events which generate a trigger, irrespective of impact parameter. 

\section{First level trigger in Ne+Ne collisions}
\label{sec:fnene}

\subsection{Minimum total multiplicity}

Due to the low multiplicity of the Ne+Ne system, a substantial fraction of 
central events has no charged tracks in one or more of the azimuthal sectors. 
Hence even the lowest sector multiplicity condition of one particle per 
sector results in a low efficiency for Ne+Ne events. In order to define the 
trigger transition, the sector multiplicity condition used in the Au+Au system 
is therefore replaced by a minimal total multiplicity condition M$_{L}$.
Fig. \ref{fig:fig7} shows the trigger efficiency for Ne+Ne events as a 
function  of the total multiplicity condition M$_{L}$. Shown are the data 
points for events with impact parameters of 1,2 and 3 fm, respectively.
For all the impact parameters shown in Fig. \ref{fig:fig7}, the efficiency 
exhibits a plateau of nearly 100\% at low total multiplicity values but drops 
steeply at large values. The optimal choice for the low total 
multiplicity condition is a value as large as possible but still within 
the plateau of the impact parameter b\,=\,1\,fm. Hence, the low multiplicity 
condition is set to M$_{L} \geq 6$ for the calculations shown below.

\subsection{Trigger timing}

Fig. \ref{fig:fig8} shows the timing of the trigger transition for events 
with different impact parameters. Here, the trigger transition is defined by 
the condition of minimum total multiplicity M$_{L} \geq 6$. Events with 
impact parameter b\,=\,1\,fm are represented by 
the solid line. The FWHM of their time distribution amounts to about 2 nsec.
The FWHM value for events with impact parameters of 
2 and 3 fm  are about 3 and 4 nsec, respectively. 
Semi-central events meet the trigger condition only with the help of
slower moving particles. These slow particles result in the asymmetric tail 
of the time distribution seen in Fig. \ref{fig:fig8}.

\subsection{Total multiplicity}

The low total multiplicity requirement M$_{L}$ defines a condition on minimum
total particle multiplicity. As in the Au+Au system, an additional condition on 
maximum event multiplicity developing within a time window following the trigger
transition might further reduce non-central events. Fig. \ref{fig:fig9} 
displays the maximum total multiplicity M$_{H}$ reached during a time window 
of 20 nsec following the trigger transition. Here, events with impact 
parameters b=1,2 and 3 fm develop different total event multiplicities. An 
additional condition on maximum event multiplicity reduces semi-central 
events considerably while affecting central events only little.

\subsection{Timing total multiplicity}

The time at which the maximum event multiplicity M$_{H}$ is reached needs to 
be known if an additional condition on total multiplicity M$_{H}$ is to be 
applied. Here, the time zero is the time of the trigger transition defined by 
the low total multiplicity condition M$_{L}$. Central and non-central events 
develop their maximum total multiplicity within a time span of 15 nsec 
following the trigger transition. Hence, similarly to the Au+Au system, a time 
window of 15-20 nsec after the trigger transition seems adequate to test for 
the maximum total event multiplicity. 

\subsection{Trigger efficiency}

In the Ne+Ne system, deadtime losses are treated in the same way as in the
system Au+Au. Fig. \ref{fig:fig10} shows the values Eff$_{LV1} 
\times$ R$_{DT}$ for central events as a function of the required total 
multiplicity M$_{H}$. The data points for zero deadtime represent the first 
level trigger efficiency. For the deadtime of 10 $\mu$sec, the Eff$_{LV1} 
\times$ R$_{DT}$ value has a maximum at an M$_{H}$ value of about 11. For the 
calculations shown below, the total multiplicity condition is therefore set 
to M$_{H} \geq 11$. At lower M$_{H}$ 
values, semi-central events do not get suppressed efficiently. Hence 
the increased rate of semi-central events leads to a decrease of the  
Eff$_{LV1} \times$ R$_{DT}$ values of central events. At high
M$_{H}$ values, Eff$_{LV1} \times$ R$_{DT}$ decreases since central
events are rejected due to insufficient multiplicity.
The data for the deadtime of 6 $\mu$sec are shown for comparison.   

\subsection{Trigger rate}

Table \ref{tab:tab3} lists the trigger rates for the Ne+Ne system. 
These rates are shown in units of 10$^{5}$ per second for the different total 
multiplicity thresholds M$_{H}$ and for deadtimes of T$_{0}$ = 0,6 and 10 
$\mu$sec, respectively. A first level trigger rate of 1.$\times$10$^{5}$ 
per second is the input design value for the next trigger stage. 
Each T$_{0}$ value in Table \ref{tab:tab3} corresponds to two columns. The 
left column shows the rate of trigger transitions resulting from the low 
total multiplicity condition M$_{L} \geq 6$. The right column displays 
the rate if additionally the total multiplicity condition M$_{H}$ is required. 

Table \ref{tab:tab4} lists the partial trigger rates from events of impact 
parameters b = 1,2 and 3 fm. Every impact parameter corresponds to two columns. 
On the left, the partial trigger rate is shown in units of 10$^{5}$ per second. 
On the right, the partial rates are normalized to the total trigger rate and 
shown in units of percent.

\subsection{First level trigger based on M$_{H}$}

As in the Au+Au system, a direct one step trigger scheme is also feasible in the
Ne+Ne system. Fig. \ref{fig:fig11} shows the trigger timing achieved by the 
one and two step trigger based on multiplicity conditions of M$_{L} \geq 11$ and 
M$_{L} \geq 6$, M$_{H} \geq 11$, respectively. The solid line shows the timing 
in the two step scheme as explained above. These data points correspond to Fig. 
\ref{fig:fig8} with the additional condition of total multiplicity M$_{H} \geq
11$. The FWHM of this distribution is about 2 nsec. The dashed line in Fig. 
\ref{fig:fig11} represents the timing in the one step trigger. Here, the total 
multiplicity condition M$_{L} \geq 11$ defines the trigger. As 
expected, the timing is delayed compared to the two step trigger and considerably 
broadened. This distribution has a FWHM of about 6 nsec and an asymmetric tail
at the high value side. The data shown in Fig. \ref{fig:fig11} represent the 
timing of all events which generate a trigger, irrespective of impact parameter. 

\section{Conclusions}

Simulations of the HADES first level trigger in both heavy-ion systems Au+Au 
and Ne+Ne indicate that the first level trigger can be implemented in both a 
one or two step trigger architecture. While both approaches yield the same 
trigger rate, the two step scheme results in a considerably improved timing 
of the trigger signal derived from the multiplicity conditions imposed on the 
TOF paddles. The two step scheme requires different multiplicity definitions 
for establishing the trigger transition in the two systems studied. A sector 
multiplicity condition is applied in the high multiplicity system Au+Au, 
but a requirement on total multiplicity is needed in the low multiplicity  
system Ne+Ne. By judicious choice of multiplicity specification for asserting 
the trigger transition, the number of central collisions passed onto the 
next trigger stage can be maximized. The trigger rate resulting from such 
multiplicity provisions satisfies the second level trigger requirement of 
10$^{5}$ events per second in both the Au+Au and Ne+Ne systems.

\begin{ack}

The support of the lepton group at GSI and, in particular, fruitful 
discussions with W.Koenig are gratefully acknowledged. The authors thank 
Gy.Wolf for providing the BUU data files used in the simulations.

\end{ack}

\newpage
{\Large \bf
\noindent
Figure Captions
}

\begin{figure}[ht]
\vspace{1.3cm}
\caption{                 
First level trigger efficiency for the system Au+Au at 1 AGeV
as a function of the sector multiplicity condition M$_{S}$.
Shown are data points for events with impact parameters of b=1,3 and 5\,fm.
}
\label{fig:fig1}
\end{figure}

\begin{figure}[ht]
\vspace{1.3cm}
\caption{                 
Trigger timing of first level trigger signal for impact parameters
b=1,3 and 5\,fm. The time zero is the time of reaction.
}
\label{fig:fig2}
\end{figure}

\begin{figure}[ht]
\vspace{1.3cm}
\caption{                 
The maximum total event multiplicity for impact parameters b=1,3 and 5\,fm 
reached during a time window of 20\,nsec following the trigger transition.
}
\label{fig:fig3}
\end{figure}

\begin{figure}[ht]
\vspace{1.3cm}
\caption{                 
Time at which the maximum total event multiplicity is reached for impact 
parameters b=1,3 and 5\,fm. The time zero is the time of trigger transition.
}
\label{fig:fig4}
\end{figure}

\begin{figure}[ht]
\vspace{1.3cm}
\caption{                 
Efficiency of first level trigger (deadtime losses included) for central
events. Shown are the data points as a function of the total multiplicity 
condition M$_{T}$ for deadtimes of 0,6 and 10\,$\mu$sec.
}
\label{fig:fig5}
\end{figure}

\begin{figure}[ht]
\vspace{1.3cm}
\caption{                 
Trigger timing in the one (dashed line) and two step (solid line) trigger 
scheme (see text). The time zero is the time of reaction.
}
\label{fig:fig6}
\end{figure}

\begin{figure}[ht]
\vspace{1.3cm}
\caption{                 
First level trigger efficiency for the system Ne+Ne at 2 AGeV
as a function of the minimum total multiplicity M$_{L}$.
Shown are data points for events with impact parameters of b=1,2 and 3\,fm.
}
\label{fig:fig7}
\end{figure}

\begin{figure}[ht]
\vspace{1.3cm}
\caption{                 
Trigger timing of first level trigger signal for impact parameters
b=1,2 and 3\,fm. The time zero is the time of reaction.
}
\label{fig:fig8}
\end{figure}

\begin{figure}[ht]
\vspace{1.3cm}
\caption{                 
The maximum total event multiplicity for impact parameters b=1,2 and 3\,fm 
reached during a time window of 20\,nsec following the trigger transition.
}
\label{fig:fig9}
\end{figure}

\begin{figure}[ht]
\vspace{1.3cm}
\caption{                 
Efficiency of first level trigger (deadtime losses included) for central
events. Shown are the data points as a function of the total multiplicity 
condition M$_{H}$ for deadtimes of 0,6 and 10\,$\mu$sec.
}
\label{fig:fig10}
\end{figure}

\begin{figure}[ht]
\vspace{1.3cm}
\caption{                 
Trigger timing in the one (dashed line) and two step (solid line) trigger 
scheme (see text). The time zero is the time of reaction.
}
\label{fig:fig11}
\end{figure}

\clearpage

\vspace{0.cm}
{\Large \bf
\noindent
Tables
}

\vspace{1.cm}
\begin{table}[ht]
\begin{tabular}{||c||c|c||c|c||c|c||} \hline
Au + Au&\multicolumn{2}{c||}{T$_{0}$\,=\,0\,$\mu$sec } 
&\multicolumn{2}{c||}{T$_{0}$\,=\,6\,$\mu$sec } 
&\multicolumn{2}{c||}{T$_{0}$\,=\,10\,$\mu$sec } \\ \cline{2-7}
&\multicolumn{2}{c||}{rate [10$^{5}$]} 
&\multicolumn{2}{c||}{rate [10$^{5}$]} 
&\multicolumn{2}{c||}{rate [10$^{5}$]} \\ \cline{1-7}
$M_{T} \geq 100$ & 1.08 & .371 & .896 & .308 & .794 & .276 \\ 
$M_{T} \geq 105$ & 1.08 & .277 & .939 & .242 & .853 & .222 \\ 
$M_{T} \geq 110$ & 1.08 & .189 & .980 & .172 & .914 & .162 \\ 
$M_{T} \geq 115$ & 1.08 & .133 & 1.01 & .124 & .965 & .118 \\ 
$M_{T} \geq 120$ & 1.09 & .081 & 1.04 & .077 & 1.01 & .075 \\ \hline
\end{tabular}
\vspace{1.cm}
\caption{
Rates of first level trigger transitions (left column) and of triggers 
(right column) in the system Au+Au (see text). The rates are shown in 
units of 10$^{5}$ per second for deadtimes T$_{0}$ = 0,6 and 10\,$\mu$sec 
and for different total multiplicity conditions M$_{T}$. 
\label{tab:tab1}
}
\end{table}

\vspace{2.cm}
\begin{table}[ht]
\begin{tabular}{||c||c|c||c|c||} \hline
Au + Au&\multicolumn{2}{c||}{rate b=1fm } 
&\multicolumn{2}{c||}{rate b\,=\,3\,fm} \\ \cline{2-5}
& [10$^{5}$] & [\%] & [10$^{5}$] & [\%] \\ \cline{1-5} 
$M_{T} \geq 100$ & .110 & 39.7 & .165 & 59.8 \\ 
$M_{T} \geq 105$ & .113 & 51.0 & .108 & 48.6 \\ 
$M_{T} \geq 110$ & .113 & 69.5 & .048 & 29.9 \\ 
$M_{T} \geq 115$ & .095 & 80.3 & .023 & 19.0 \\ 
$M_{T} \geq 120$ & .068 & 90.3 & .006 & 8.7 \\ \hline
\end{tabular}
\vspace{1.cm}
\caption{
First level partial trigger rates of events with impact parameters b\,=\,1 and 
3\,fm in the system Au+Au for different total multiplicity conditions M$_{T}$.
The left column displays the partial rate in units of 10$^{5}$ per second. In 
the right column, the partial rates are normalized to the total trigger rate 
and shown in units of percent.
\label{tab:tab2}
}
\end{table}
\vspace{2.cm}
\begin{table}[ht]
\begin{tabular}{||c||c|c||c|c||c|c||} \hline
Ne + Ne&\multicolumn{2}{c||}{T$_{0}$\,=\,0\,$\mu$sec } 
&\multicolumn{2}{c||}{T$_{0}$\,=\,6\,$\mu$sec } 
&\multicolumn{2}{c||}{T$_{0}$\,=\,10\,$\mu$sec } \\ \cline{2-7}
&\multicolumn{2}{c||}{rate [10$^{5}$]} 
&\multicolumn{2}{c||}{rate [10$^{5}$]} 
&\multicolumn{2}{c||}{rate [10$^{5}$]} \\ \cline{1-7}
$M_{H} \geq 7$ & 2.20 & 1.85 & 1.05 & .884 & .774 & .651 \\ 
$M_{H} \geq 9$ & 2.21 & 1.30 & 1.26 & .735 & .966 & .567 \\ 
$M_{H} \geq 11$ & 2.22 & .865 & 1.47 & .574 & 1.19 & .464 \\ 
$M_{H} \geq 13$ & 2.22 & .532 & 1.69 & .408 & 1.45 & .350 \\ 
$M_{H} \geq 15$ & 2.22 & .290 & 1.90 & .248 & 1.73 & .226 \\ \hline
\end{tabular}
\vspace{1.cm}
\caption{
Rates of first level trigger transitions (left column) and of triggers 
(right column) in the system Ne+Ne (see text). The rates are shown in 
units of 10$^{5}$ per second for deadtimes T$_{0}$ = 0,6 and 10\,$\mu$sec 
and for different total multiplicity conditions M$_{H}$. 
\label{tab:tab3}
}
\end{table}

\vspace{.6cm}
\begin{table}[ht]
\begin{tabular}{||c||c|c||c|c||c|c||} \hline
Ne + Ne&\multicolumn{2}{c||}{rate b=1fm } 
&\multicolumn{2}{c||}{rate b\,=\,2\,fm} 
&\multicolumn{2}{c||}{rate b\,=\,3\,fm} \\ \cline{2-7}
& [10$^{5}$] & [\%] & [10$^{5}$] & [\%] & [10$^{5}$] & [\%] \\ \cline{1-7} 
$M_{H} \geq 7$ & .135 & 20.7 & .215 & 33.0 & .209 & 32.0 \\ 
$M_{H} \geq 9$ & .155 & 27.3 & .225 & 39.6 & .149 & 26.3 \\ 
$M_{H} \geq 11$ & .164 & 35.4 & .197 & 42.5 & .089 & 19.2 \\ 
$M_{H} \geq 13$ & .159 & 45.4 & .141 & 40.2 & .046 & 13.0 \\ 
$M_{H} \geq 15$ & .122 & 54.2 & .082 & 36.4 & .019 & 8.3 \\ \hline
\end{tabular}
\vspace{1.cm}
\caption{
First level partial trigger rates of events with impact parameters b\,=\,1,\,2 
and 3\,fm in the system Ne+Ne for different total multiplicity conditions 
M$_{H}$. The left column displays the partial rate in units of 10$^{5}$ per 
second. In the right column, the partial rates are normalized to the total 
trigger rate and shown in units of percent.
\label{tab:tab4}
}

\end{table}

\vspace*{5cm}

\newpage

\pagestyle{empty}

\begin{minipage}{16.cm}
\epsfig{figure=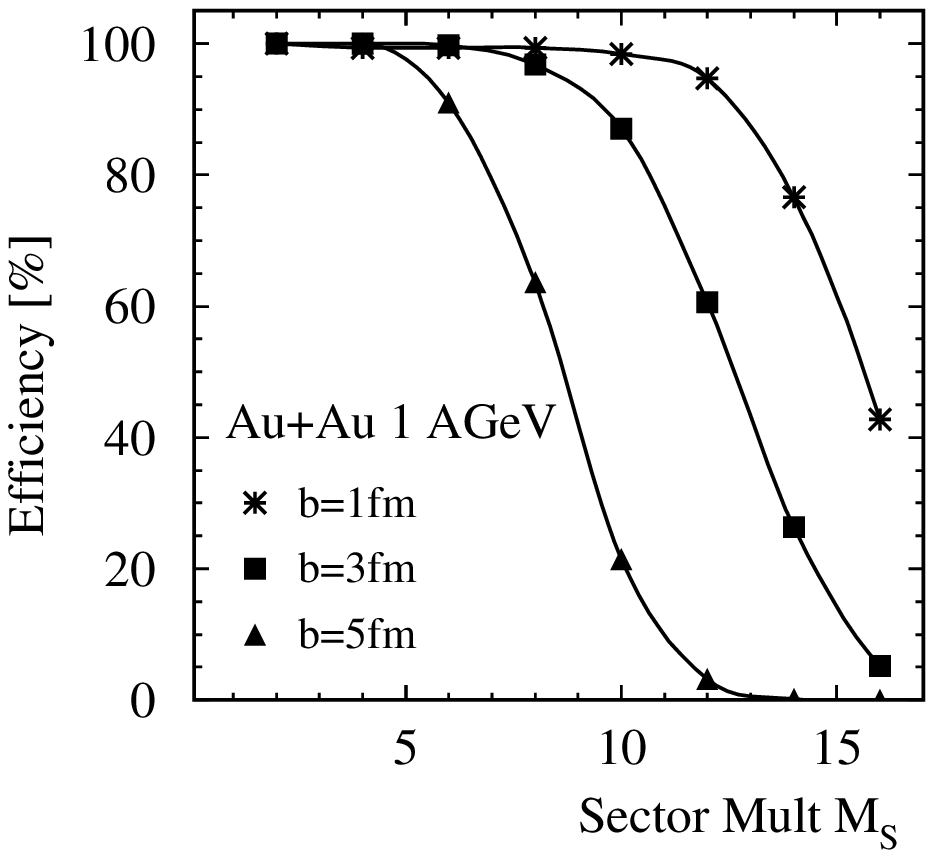,width=.85\linewidth}
\hfill
\end{minipage}

\begin{minipage}{2.cm}
\vspace{3.cm}
{\bf\huge FIG.1}
\end{minipage}

\newpage

\begin{minipage}{15.cm}
\epsfig{figure=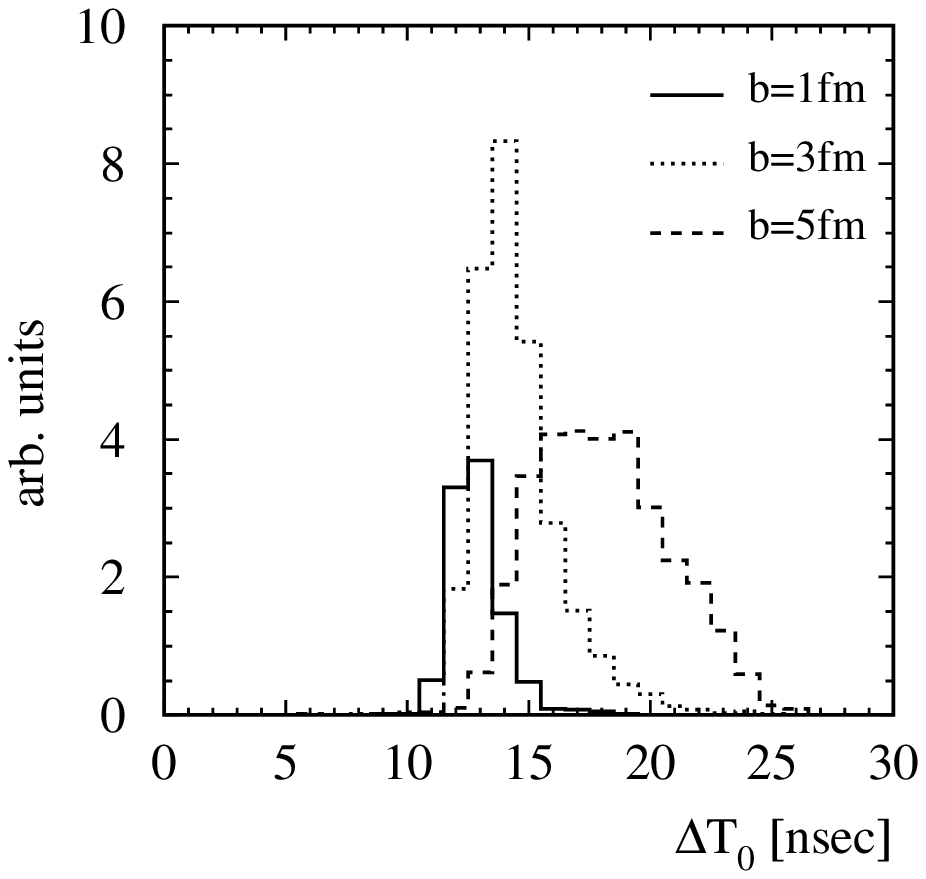,width=.99\linewidth}
\hfill
\end{minipage}

\begin{minipage}{2.cm}
\vspace{3.cm}
{\bf\huge FIG.2}
\end{minipage}

\newpage

\begin{minipage}{15.cm}
\epsfig{figure=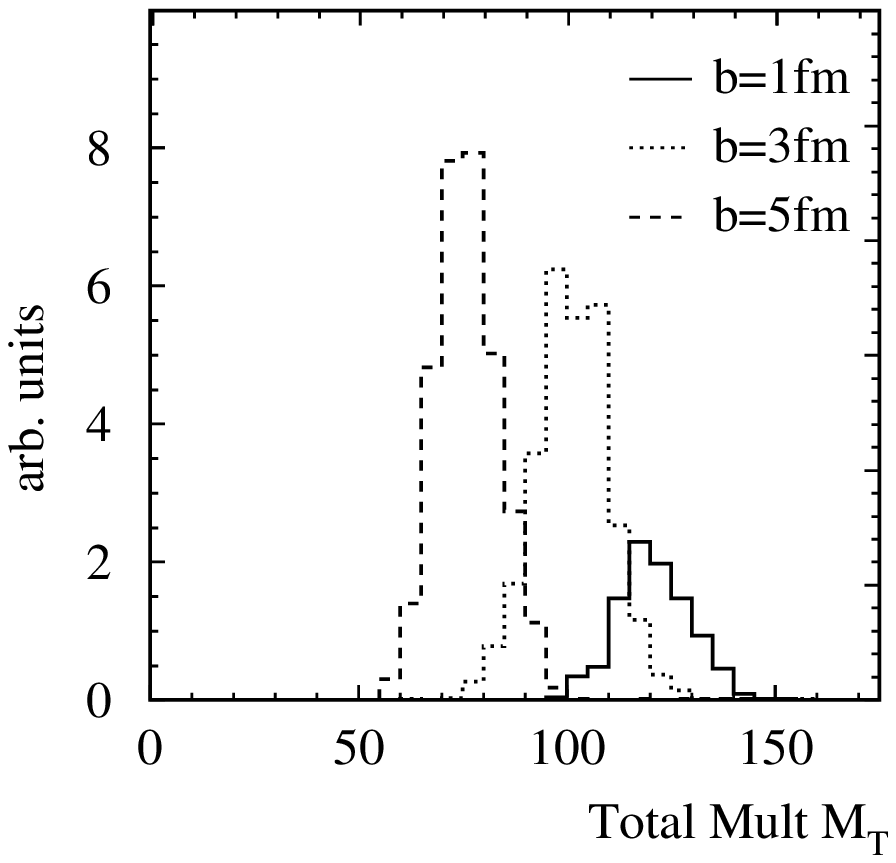,width=.99\linewidth}
\hfill
\end{minipage}

\begin{minipage}{2.cm}
\vspace{3.cm}
{\bf\huge FIG.3}
\end{minipage}

\newpage

\begin{minipage}{15.cm}
\epsfig{figure=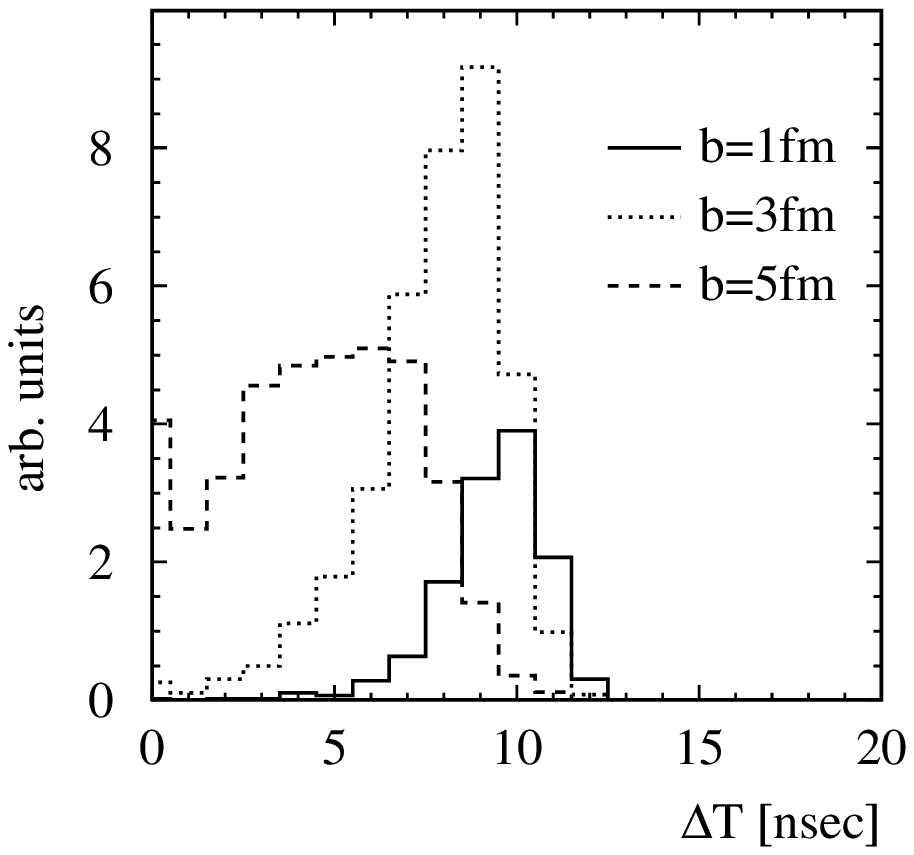,width=.99\linewidth}
\hfill
\end{minipage}

\begin{minipage}{2.cm}
\vspace{3.cm}
{\bf\huge FIG.4}
\end{minipage}

\newpage

\begin{minipage}{15.cm}
\epsfig{figure=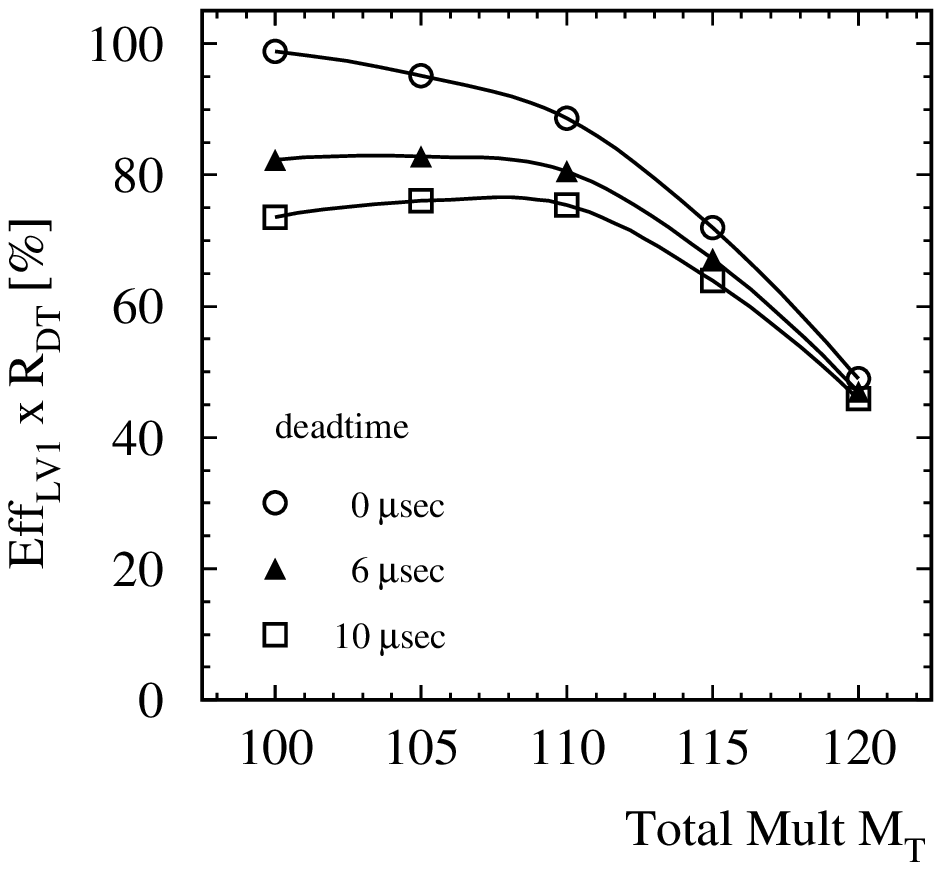,width=.99\linewidth}
\hfill
\end{minipage}

\begin{minipage}{2.cm}
\vspace{3.cm}
{\bf\huge FIG.5}
\end{minipage}

\newpage

\begin{minipage}{15.cm}
\epsfig{figure=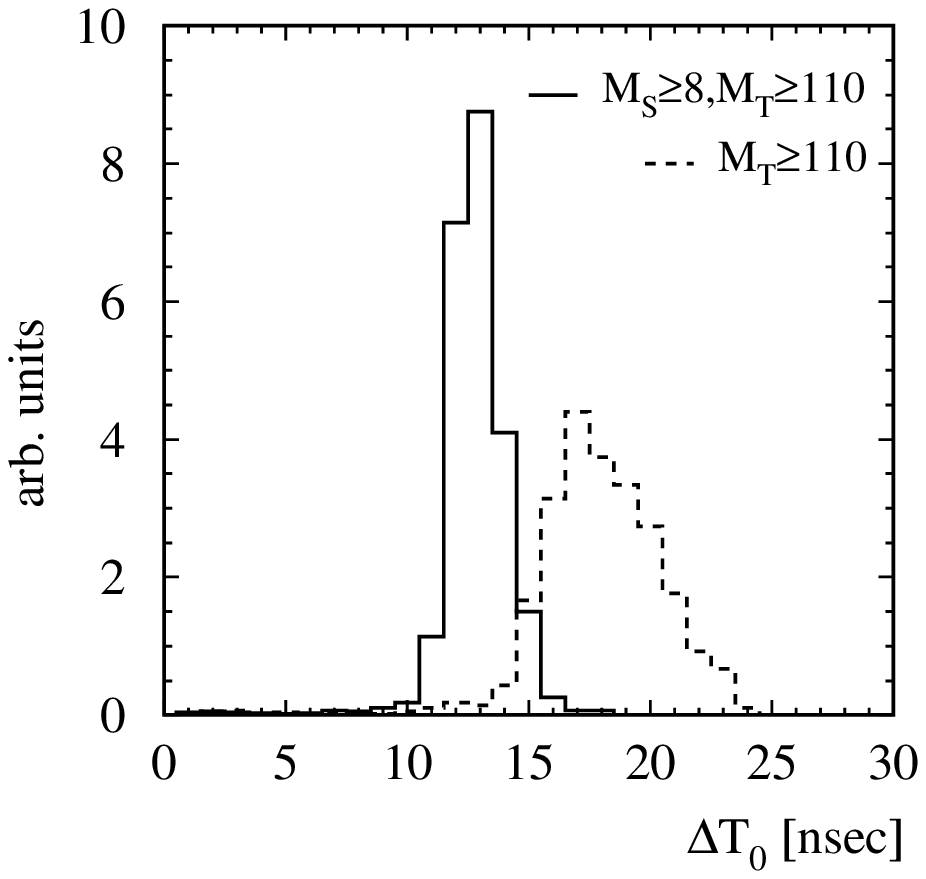,width=.99\linewidth}
\hfill
\end{minipage}

\begin{minipage}{2.cm}
\vspace{3.cm}
{\bf\huge FIG.6}
\end{minipage}

\newpage

\begin{minipage}{15.cm}
\epsfig{figure=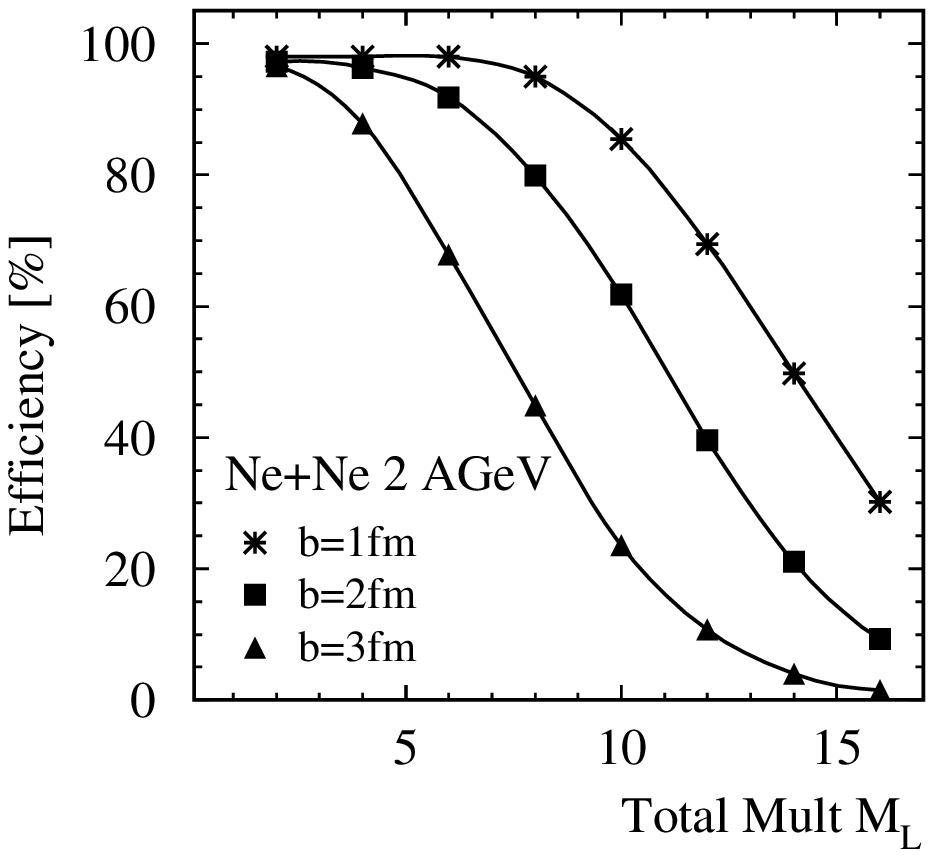,width=.99\linewidth}
\hfill
\end{minipage}

\begin{minipage}{2.cm}
\vspace{3.cm}
{\bf\huge FIG.7}
\end{minipage}

\newpage

\begin{minipage}{15.cm}
\epsfig{figure=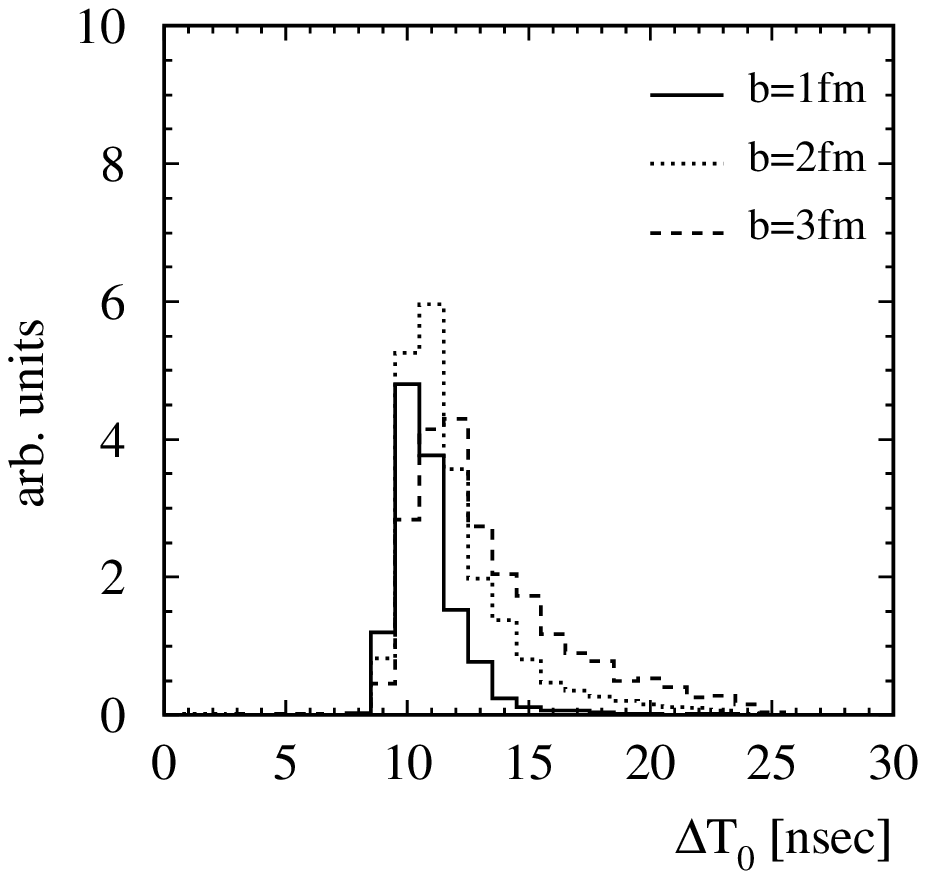,width=.99\linewidth}
\hfill
\end{minipage}

\begin{minipage}{2.cm}
\vspace{3.cm}
{\bf\huge FIG.8}
\end{minipage}

\newpage

\begin{minipage}{15.cm}
\epsfig{figure=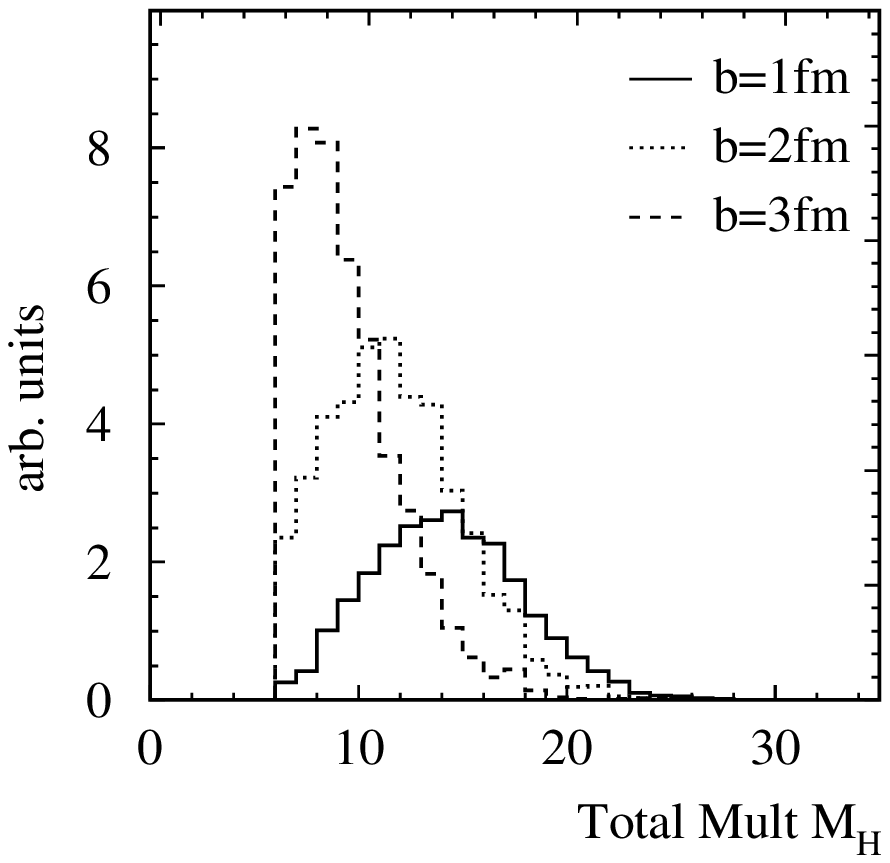,width=.99\linewidth}
\hfill
\end{minipage}

\begin{minipage}{2.cm}
\vspace{3.cm}
{\bf\huge FIG.9}
\end{minipage}

\newpage

\begin{minipage}{15.cm}
\epsfig{figure=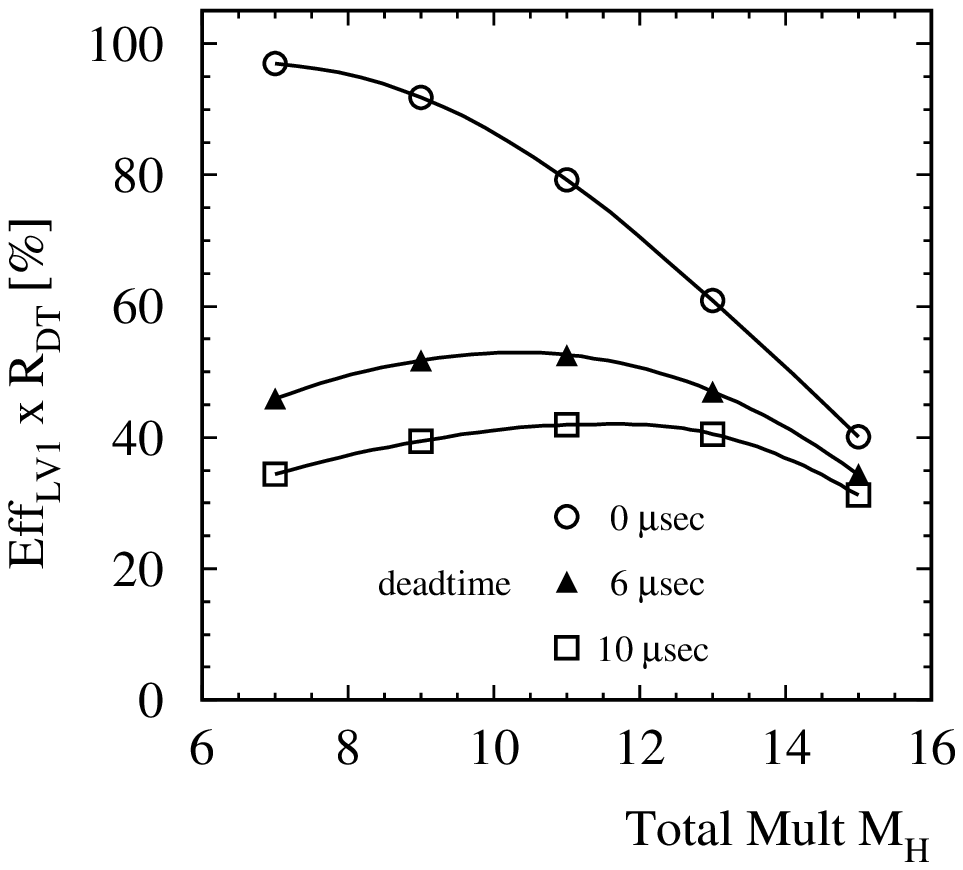,width=.99\linewidth}
\hfill
\end{minipage}

\begin{minipage}{2.cm}
\vspace{3.cm}
{\bf\huge FIG.10}
\end{minipage}

\newpage

\begin{minipage}{15.cm}
\epsfig{figure=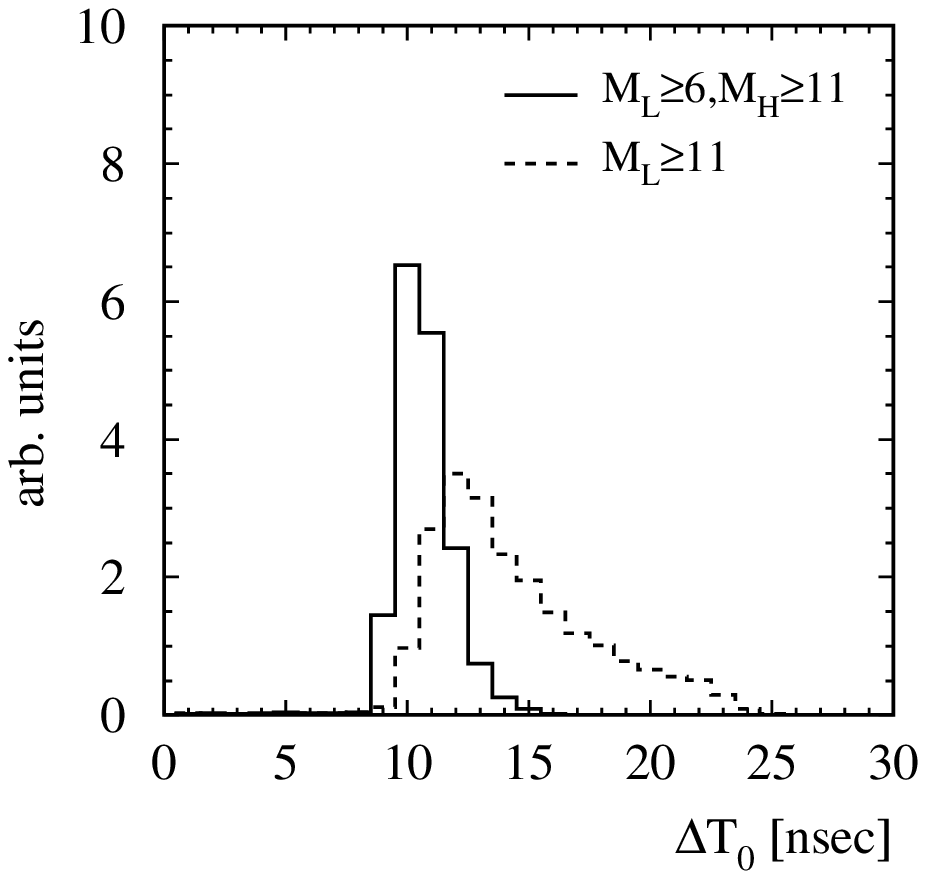,width=.99\linewidth}
\hfill
\end{minipage}

\begin{minipage}{2.cm}
\vspace{3.cm}
{\bf\huge FIG.11}
\end{minipage}

\end{document}